# On the Reliability of Funding Acknowledgements as Research Data: Evidence from Astronomy

**Stahlman, Gretchen**  Rutgers University, USA | gretchen.stahlman@rutgers.edu

## ABSTRACT
Online bibliographic databases have enabled new research through which bibliographic records are analyzed as data about science. Within these records, the acknowledgements sections of papers are often used to draw conclusions about funding support for published research. While acknowledgements and funding statements can be informative for research and policy development, this poster adds to a body of literature that highlights limitations of funding data for scientometric and policy research, using evidence gathered from a questionnaire of authors of astronomy journal articles. The study shows that only 71.4% of a sample of authors of papers tied to NSF grants through acknowledgements reported in the survey that NSF funded the research presented in the respective papers. A brief analysis of the questionnaire followed by recommendations and considerations for further research are presented. The discrepancy in reporting appears to indicate that funding streams can be fluid and not always apparent to authors, overall raising the question of what sorts of research should be addressed with funding statements, where conceptually tying a paper directly to a grant is not straightforward.

## KEYWORDS
research funding, scholarly communication, acknowledgements, scientometrics, astronomy

## INTRODUCTION
The evolution of online bibliographic databases has enabled scientometric and policy-oriented research, where bibliographic records can be analyzed as data. Within these records, the acknowledgements sections of papers are often used to draw conclusions about research support and conflicts of interest, indirectly influencing the evolution of research policy and funding programs across disciplines. Authors typically mention funding agencies and grant numbers within acknowledgements, leading to efforts to connect specific grants and amounts of funding directly to papers to evaluate research outcomes and return on investment. A few examples of this type of research include Kurczynski & Milojevic (2020), Larivière & Sugimoto (2018), and Tatsioni, Vavva & Ioannidis (2010). While acknowledgements and funding statements can be informative, this poster explores potential limitations of using funding data for research and policy through evidence gathered from a survey of authors of astronomy journal articles.

## BACKGROUND
Scholars have identified limitations of using funding acknowledgements to draw conclusions about research practices. For example, Rigby (2011) shows that there is a tendency to exaggerate productivity of certain grants. Paul-Hus, Desrochers and Costas (2016) note that "funding acknowledgement data remain self-declared information and are thus subject to unethical or inconsistent behaviours, either when authors fail to acknowledge funding sources or when, on the contrary, they acknowledge support they did not actually receive" (p. 3). As machine learning and natural language processing techniques are increasingly used to process funding information corresponding to papers, it is necessary to consider the types of research questions and methods that can benefit from funding acknowledgements. Through a dissertation study documented in Stahlman (2020) and Stahlman & Heidorn (2020), a series of incidental findings is presented here, adding insight to conversations about research with funding statements.

## METHODS AND RESULTS
Using author email addresses obtained from Web of Science bibliographic records, a survey was sent to corresponding authors of astronomy papers acknowledging a sample of NSF Division of Astronomical Sciences (AST) grant numbers. The sampled grants originated in 2016, and the survey was conducted between May and June 2019. The survey was sent to 477 authors of papers associated with sampled grants, and 107 responses were received corresponding to papers published between 2016 and 2019 in 13 astronomy journals. The original purpose of the survey was to obtain information about the locations and characteristics of astronomical data that correspond to the papers, and the same survey was also sent to a second population not discussed here. As part of this larger overarching study (Stahlman, 2020), participants associated with the known sample of NSF grants through acknowledgements in the literature were nevertheless asked the multiple-choice question: *Which agency or agencies funded the research presented in this paper? Select all that apply.* Options for responses were: 1) *NASA*, 2) *NSF*, 3) *DOE*, 4) *DOD*, 5) *Institutional or university support*, 6) *Private foundation(s)*, 7) *International (Non-U.S.) agency*, 8) *Other (specify)*, and 9) *Not applicable*. Surprisingly, only 71.4% of the authors of papers linked to NSF grants through Web of Science indicated NSF as a funder of the research in the respective question (n=107).

Review of the funding statements eliminated the possibility of "false positives" originating with the Web of Science search, as all acknowledgements referred to NSF grants in some way. However, a variety of acknowledgement styles appeared across both the group that indicated NSF funding on the questionnaire and the group that did not. For example, some funding statements are quite long and detailed and others more concise. Also, some funding statements use language clearly indicating direct support of the research at hand by NSF and others indicate piecemeal individual support held by each author. Overall, a clear pattern was not detected through manual review alone.

Another possible explanation for the discrepancy is that the corresponding author who completed the survey was not immediately cognizant of all grants held by coauthors. In astronomy and other fields, the corresponding author designation is a respected role with special responsibility, and this individual is expected to be highly familiar with the research. Especially considering the freshness of sampled grants and papers at the time the questionnaire was completed, it is difficult to dismiss outright the impressions of the corresponding authors about which major funding agencies supported their papers. A binary variable was created to indicate whether a respondent from the NSF-acknowledged sample also selected NSF as a funder in response to the survey question, and significance tests were conducted (with 95% confidence interval) to further deconstruct authorship characteristics that may contribute to the discrepancy, reported briefly below.

*H1: The discrepancy in reporting the funding for a paper is related to the number of authors on the paper.* If a paper has many authors, it may be more difficult for the corresponding author to be aware of funding held by co-authors. The number of authors on sampled papers ranged from 2 to 122; a new log transformed variable was created and a Welch's two sample t-test was conducted. The result was not significant at the .05 level but may be considered marginally significant and worthy of further exploration ($t = 1.8333$, $df = 58.82$, p-value = 0.07181).

*H2: The discrepancy in reporting the funding for a paper is related to international collaboration.* If a paper involves international collaboration between U.S. and non-U.S. authors, it may be more difficult for the corresponding author to be aware of funding held by coauthors. A Pearson's Chi-squared test was conducted, and the result was not significant (chi-squared = 2.5885, $df = 1$, p-value = 0.1076).

*H3: The discrepancy in reporting the funding for a paper is related to the career stage of the corresponding author.* If the corresponding author is very early in their career, they may be less familiar with funding reporting practices and other norms. To determine whether career stage of the corresponding author is significantly related to the discrepancy, a Welch's two sample t-test was conducted, and the result was not significant ($t = 0.74725$, $df = 65.841$, p-value = 0.4576).

*H4: The discrepancy in reporting the funding for a paper is related to whether the corresponding author is also the first author.* For the present study, a majority of corresponding author respondents - 100 out of 107 - are first authors as well, which does not support statistical inference. It may be worth noting that four out of the seven non-first-authors fall into the discrepancy category (57%). This percentage is higher than the larger dataset (28.6%), but there are too few observations to draw a conclusion.

**DISCUSSION AND CONCLUSION**

The incidental findings and brief exploration presented here demonstrate that funding streams in astronomy are fluid and not always apparent to authors. This conclusion aligns with the argument of Rigby (2011) that links between papers and acknowledged funding are complicated and often indirect, where research funding essentially supports an ecosystem of processes rather than specific papers. By directly obtaining authors' perspectives through a questionnaire about specific papers, the present study has illuminated discrepancies in awareness and reporting of funding sources. That said, possible reasons were identified for the discrepancy – the most promising explanation (within the limitations of the study) being that having more authors on a paper may contribute to enhanced ambiguity about research support. Features of astronomical research are unique and could contribute to the discrepancy as well, as authors may or may not choose to acknowledge the grant numbers of NSF-funded facilities and instruments, and where investments in primary research are balanced with research support-related funding for training, major facilities, instrumentation, and software (Stahlman & Heidorn, 2020).

Beyond the small study in one discipline presented here, the overall complexity of funding acknowledgements demonstrated in the broader literature warrants continued caution when using funding statements to generalize. This issue raises the question of what sorts of research can or should be addressed with funding statements, where conceptually tying papers directly to grants is not straightforward. Future work will further explore of the quality and integrity of acknowledgements as a data source, to avoid natural language pitfalls with automatic extraction and ensure accurate reporting of research outputs to justify funding and assess the overall value and impact of scientific research. These issues also point to a need for further qualitative research on the nuances of acknowledgement behavior and impressions of authors across disciplines about direct and indirect funding support for papers.

**REFERENCES**

Kurczynski, P. L., & Milojevic, S. (2020). Enabling discoveries: A review of 30 years of advanced technologies and instrumentation at the National Science Foundation. *Journal of Astronomical Telescopes, Instruments, and Systems*, *6*(3), 030901. https://doi.org/10.1117/1.JATIS.6.3.030901

Larivière, V., & Sugimoto, C. R. (2018). Do authors comply when funders enforce open access to research? *Nature*, *562*(7728), 483–486. https://doi.org/10.1038/d41586-018-07101-w

Paul-Hus, A., Desrochers, N., & Costas, R. (2016). Characterization, description, and considerations for the use of funding acknowledgement data in Web of Science. *Scientometrics*, *108*(1), 167–182. https://doi.org/10.1007/s11192-016-1953-y

Rigby, J. (2011). Systematic grant and funding body acknowledgement data for publications: New dimensions and new controversies for research policy and evaluation. *Research Evaluation*, *20*(5), 365–375. https://doi.org/10.3152/095820211X13164389670392

Stahlman, G. R. (2020). *Exploring the long tail of astronomy: A mixed-methods approach to searching for dark data* (Doctoral dissertation, The University of Arizona). https://www.proquest.com/docview/2435763825/abstract/8E65945F493F45AAPQ/1

Stahlman, G. R., & Heidorn, P. B. (2020). Mapping the "long tail" of research funding: A topic analysis of NSF grant proposals in the division of astronomical sciences. *Proceedings of the Association for Information Science and Technology*, *57*(1), e276. https://doi.org/10.1002/pra2.276

Tatsioni, A., Vavva, E., & Ioannidis, J. P. A. (2010). Sources of funding for Nobel Prize-winning work: Public or private? *The FASEB Journal*, *24*(5), 1335–1339. https://doi.org/10.1096/fj.09-148239